\documentclass[aip,graphicx,reprint,superscriptaddress]{revtex4-1}

\usepackage[utf8]{inputenc}
\usepackage[T1]{fontenc}
\usepackage{mathptmx}
\usepackage{etoolbox}
\usepackage{graphicx}
\usepackage{amsmath}
\usepackage{amssymb}
\usepackage{hyperref}
\usepackage{multirow}
\usepackage{xcolor}

\begin{document}

\title{Revisiting N\'{e}el 60 years on: the magnetic anisotropy of L1\textsubscript{0} FeNi (tetrataenite)}

\author{Christopher D. Woodgate}
\email[]{Christopher.Woodgate@warwick.ac.uk}
\affiliation{Department of Physics, University of Warwick, Coventry, CV4 7AL, United Kingdom}
\author{Christopher E. Patrick}
\affiliation{Department of Materials, University of Oxford, Oxford, OX1 3PH, United Kingdom}
\author{Laura H. Lewis}
\affiliation{Department of Chemical Engineering, Northeastern University, Boston, MA 02115, USA}
\affiliation{Department of Mechanical and Industrial Engineering, Northeastern University, Boston, MA 02115, USA}
\author{Julie B. Staunton}
\email[]{J.B.Staunton@warwick.ac.uk}
\affiliation{Department of Physics, University of Warwick, Coventry, CV4 7AL, United Kingdom}

\date{\today}

\begin{abstract}
The magnetocrystalline anisotropy energy of atomically ordered L1\textsubscript{0} FeNi (the meteoritic mineral tetrataenite) is studied within a first-principles electronic structure framework.
Two compositions are examined: equiatomic Fe$_{0.5}$Ni$_{0.5}$ and an Fe-rich composition, Fe$_{0.56}$Ni$_{0.44}$. It is confirmed that, for the single crystals modelled in this work, the leading-order anisotropy coefficient $K_1$ dominates the higher-order coefficients $K_2$ and $K_3$.
To enable comparison with experiment, the effects of both imperfect atomic long-range order and finite temperature are included.
While our computational results initially appear to undershoot the measured experimental values for this system, careful scrutiny of the original analysis due to N\'{e}el {\it et al.} [J. Appl. Phys. {\bf 35}, 873 (1964)] suggests that our computed value of $K_1$ is, in fact, consistent with experimental values, and that the noted discrepancy has its origins in the nanoscale polycrystalline, multivariant nature of experimental samples, that yields much larger values of $K_2$ and $K_3$ than expected {\it a priori}.
These results provide fresh insight into the existing discrepancies in the literature regarding the value of tetrataenite's uniaxial magnetocrystalline anisotropy in both natural and synthetic samples.
\end{abstract}

\maketitle

\section{Introduction}

\begin{figure}[b]
\centering
\includegraphics[width=0.5\textwidth]{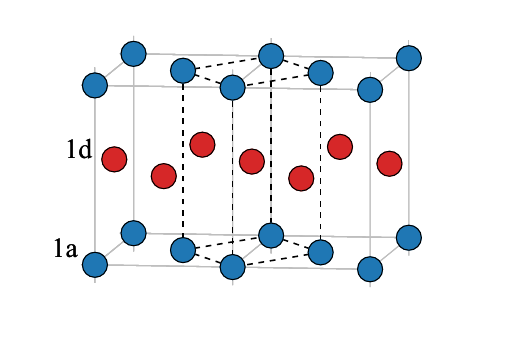}
\caption{Visualisation of the L1\textsubscript{0} ordered structure imposed on a face-centred tetragonal lattice. The conventional simple tetragonal cell, with 1a sites at the corners and a 1d site at the centre, is indicated by dashed black lines. The simple tetragonal unit cell has lattice parameters $c'=c$, $a'=b'=a/\sqrt{2}$, where $a$ and $c$ are the face-centred tetragonal lattice parameters. Perfectly ordered L1\textsubscript{0} Fe$_{0.5}$Ni$_{0.5}$ has all 1a (blue) sites occupied by Fe atoms, and all 1d (red) sites occupied by Ni atoms. The disordered Fe$_{0.5}$Ni$_{0.5}$ alloy in which every site is occupied with a 50\% probability of either Fe or Ni has a face-centred cubic lattice with $c=a$.}
\label{fig:l10_structure}
\end{figure}

Permanent magnets are of critical importance in many technological applications, prime examples being in electrical power generation and electric motors that are essential in the transition to a low-carbon future. At present, the vast majority of permanent magnets used for advanced applications contain significant amounts of rare-earth elements such as samarium and neodymium~\cite{coey_perspective_2020}, with examples being SmCo$_5$~\cite{strnat_family_1967} and Nd$_2$Fe$_{14}$B~\cite{sagawa_new_1984, croat_prfe_1984}; the latter compound also requires incorporation of rare and expensive dysprosium and/or terbium for robust high-temperature performance. These rare-earth elements are a constrained resource accompanied by concerns around price volatility and the long-term stability of the global supply chain~\cite{smith_stegen_heavy_2015, mccallum_practical_2014}. There are also concerns over the environmental impact of their extraction~\cite{bai_evaluation_2022} and processing. Overall these conditions motivate a globally concerted effort to develop rare-earth-free permanent magnets for use in advanced applications. One such candidate under consideration is atomically ordered Fe-Ni that crystallises in the tetragonal L1\textsubscript{0} structure~\cite{lewis_inspired_2014}. (The L1\textsubscript{0} structure is visualised in Fig.~\ref{fig:l10_structure}.) This compound, which is naturally found in extremely slowly cooled metallic meteorites, is also referred to as tetrataenite~\cite{clarke_tetrataeniteordered_1980}.

Although there are significant challenges associated with its formation that arise from a low atomic ordering temperature and sluggish kinetics~\cite{neel_magnetic_1964}, the L1\textsubscript{0} phase is thermodynamically stable relative to its atomically disordered, face-centered-cubic (fcc) counterpart, and a number of attempts have been made to synthesise it in the lab~\cite{neel_magnetic_1964, pauleve_magnetization_1968, shima_structure_2007, mizuguchi_artificial_2011, kojima_l1_2011, kojima_magnetic_2012, kojima_feni_2014, makino_artificially_2015, frisk_resonant_2016, frisk_strain_2017, goto_synthesis_2017, maat_creating_2020, ito_fabrication_2023, ivanov_direct_2023}. Tetrataenite is known, both from experimental measurements~\cite{neel_magnetic_1964, pauleve_magnetization_1968, shima_structure_2007, mizuguchi_artificial_2011, kojima_l1_2011, kojima_magnetic_2012, kojima_feni_2014, poirier_intrinsic_2015, makino_artificially_2015, frisk_resonant_2016, frisk_strain_2017, goto_synthesis_2017, maat_creating_2020, ito_fabrication_2023, ivanov_direct_2023} and from theoretical calculations~\cite{miura_origin_2013, edstrom_electronic_2014, lewis_magnete_2014, werwinski_ab_2017, izardar_interplay_2020, tuvshin_first-principles_2021, si_effect_2022}, to possess a sufficiently large uniaxial magnetocrystalline anisotropy energy (MAE) to be useful as a `gap magnet'~\cite{coey_permanent_2012}, with anticipated performance below that of the rare-earth magnets but above that of oxide ferrite magnets.

\begin{table*}[t]
\begin{ruledtabular}
\begin{tabular}{llllrrlll}
Reference               & Sample Type                         & Composition               & $S$ & $T$ (K) & \multicolumn{4}{l}{Anisotropy Constants (MJm$^{-3}$)} \\ \cline{6-9}
&&&&& $K_U$& $K_1$& $K_2$& $K_3$\\ \hline
N\'{e}el {\it et al.} (1964) \cite{neel_magnetic_1964}      & Bulk Sample (Disc)   & Fe$_{0.5}$Ni$_{0.5}$      & 0.41--0.45 (Estimate)                        & 293     &  0.55                  & 0.32               & 0.23               &                    \\
Paulev\'{e} {\it et al.} (1968) \cite{pauleve_magnetization_1968}   & Bulk Sample (Sphere)  & Fe$_{0.5}$Ni$_{0.5}$      &                             &         &     0.55               & 0.3                & 0.17               & 0.08               \\
Shima {\it et al.} (2007) \cite{shima_structure_2007} & Film                & Fe$_{0.5}$Ni$_{0.5}$      & 0.6$\pm$0.2                        & Room T.     & 0.63               &                    &                    &                    \\
Mizugichi {\it et al.} (2011) \cite{mizuguchi_artificial_2011} & Thin Film                & Fe$_{0.5}$Ni$_{0.5}$      & 0.33                        & Room T.     & 0.58               &                    &                    &                    \\
Kojima {\it et al.} (2011) \cite{kojima_l1_2011}    & Film                     & Fe$_{0.5}$Ni$_{0.5}$      & 0.5$\pm$0.1                 & Room T.     & 0.50$\pm$0.01        &                    &                    &                    \\
Kojima {\it et al.} (2012) \cite{kojima_magnetic_2012}    & Film                     & Fe$_{0.5}$Ni$_{0.5}$      & 0.48$\pm$0.05               & Room T.     & 0.70$\pm$0.02        &                    &                    &                    \\
Kojima et al. (2014) \cite{kojima_feni_2014}   & Film                     & Fe$_{0.5}$Ni$_{0.5}$      & 0.4                   & Room T.     & 0.7                &                    &                    &                    \\
                        &       Film                              & Fe$_{0.6}$Ni$_{0.4}$      & 0.2                   & Room T.     & 0.93               &                    &                    &                    \\
Poirier {\it et al.} (2015) \cite{poirier_intrinsic_2015}   & Meteorite                          & Fe$_{0.57}$Ni$_{0.43}$    &                            &      & 0.84               &                    &                    &                    \\
Frisk {\it et al.} (2016) \cite{frisk_resonant_2016}     & Film                     & Fe$_{0.5}$Ni$_{0.5}$      &                             & 35     & $<$0.2                &                    &                    &                    \\
Frisk {\it et al.} (2017) \cite{frisk_strain_2017}     & Film                     & Fe$_{0.5}$Ni$_{0.5}$      &                             & 300     & $\sim$0.35               &                    &                    &                    \\
Ito {\it et al.} (2023) \cite{ito_fabrication_2023}       & Film                     & Fe$_{0.5}$Ni$_{0.5}$      & 0.60                         & Room T.     & 0.55               &                    &                    &                   
\end{tabular}
\end{ruledtabular}
\caption{Experimental measurements of the magnetocrystalline anisotropy of FeNi, with sample description and atomic order parameter, $S$, where provided. 
Blank fields in the table indicate values not provided in the associated reference. Uncertainties are quoted where provided in the associated reference. $K_U$ denotes the conventional measurement of the uniaxial anisotropy, the difference in energy between a system magnetised in the $\hat{\mathbf{z}}$ direction and one magnetised in the $\hat{\mathbf{x}}$ direction. We note that the order parameter provided by N\'{e}el {\it et al.} in Ref.~\citenum{neel_magnetic_1964} is an estimate, rather than a direct measurement.}
\label{table:experimental}
\end{table*}

To date, computational modelling of tetrataenite's magnetocrystalline anisotropy has focused on evaluating its uniaxial anisotropy energy, $K_U$, which is the difference in energy between the system magnetised in the $\hat{\mathbf{z}}$ direction compared to that along the $\hat{\mathbf{x}}$ direction, rather than resolving it into its separate coefficients, {\it i.e.} $ K_U = K_1 + K_2 + K_3$. In addition, these calculations have primarily focused on the equiatomic binary composition, Fe$_{0.5}$Ni$_{0.5}$, at $T=0$ K and with perfect atomic order\cite{izardar_interplay_2020, si_effect_2022, werwinski_ab_2017, miura_origin_2013, tuvshin_first-principles_2021, edstrom_electronic_2014, lewis_magnete_2014}. However, experimental measurements of this material's magnetocrystalline anisotropy are invariably performed at finite temperature (typically room temperature) on samples with {\it imperfect} atomic order\cite{neel_magnetic_1964, pauleve_magnetization_1968, shima_structure_2007, mizuguchi_artificial_2011, kojima_l1_2011, kojima_magnetic_2012, kotsugi_structural_2014, kojima_feni_2014, makino_artificially_2015, poirier_intrinsic_2015, frisk_resonant_2016, frisk_strain_2017, ito_fabrication_2023}. (A summary of the experimental data can be found in Table~\ref{table:experimental}.) These factors are significant, because both decreasing atomic long-range order and increasing temperature reduce the magnetocrystalline anisotropy of L1\textsubscript{0} materials\cite{staunton_temperature_2004, staunton_long-range_2004, okamoto_chemical-order-dependent_2002}. 

{Computational predictions~\cite{izardar_interplay_2020, si_effect_2022, werwinski_ab_2017, miura_origin_2013, tuvshin_first-principles_2021, edstrom_electronic_2014, lewis_magnete_2014} give an MAE for L1\textsubscript{0} Fe$_{0.5}$Ni$_{0.5}$ in the range  {0.23}--0.78~MJm$^{-3}$, while experimental observations~\cite{neel_magnetic_1964, pauleve_magnetization_1968, shima_structure_2007, mizuguchi_artificial_2011, kojima_l1_2011, kojima_magnetic_2012, kojima_feni_2014, poirier_intrinsic_2015, frisk_resonant_2016, frisk_strain_2017, ito_fabrication_2023} fall in the range 0.2--0.93~MJm$^{-3}$. Although at first glance there appears to be consistency between experiment and theory, we again emphasise that theoretical calculations are performed at $T=0$~K on systems with perfect atomic order, while experimental measurements are performed at room temperature on samples with \emph{imperfect} atomic order.} Further analysis of the experimental data~\cite{neel_magnetic_1964,lewis_inspired_2014, lewis_magnete_2014} has suggested that the MAE of this material could reach the range 1.0--1.3~MJm$^{-3}$. Typically, {\it ab initio} estimates of MAE values for L1\textsubscript{0} transition-metal/noble-metal systems have been in good agreement with the corresponding experimental data, for example for FePd~\cite{staunton_temperature_2006}, FePt~\cite{staunton_temperature_2004, staunton_long-range_2004, staunton_temperature_2006, sakuma_first_1994}, and CoPt~\cite{sakuma_first_1994}, where the {\it ab initio} values for idealised systems tend to, if anything, slightly overestimate MAE values. The discrepancy between experimental and computational results for FeNi is therefore striking, and we suggest that it highlights a gap in understanding of the underlying physical origin of magnetocrystalline anisotropy in this material.

In addition, we note that { almost all} computational studies, and most experiments, do not resolve the anisotropy energy into its separate coefficients, $K_1$, $K_2$, and $K_3$. These coefficients have their origins in the symmetry of the underlying crystal structure. For a tetragonal crystal structure, $K_1$ is the leading (second-order) term in an expansion of the magnetocrystalline anisotropy energy, while $K_2$ and $K_3$ are higher-order (fourth order) terms~\cite{landau_electrodynamics_2009}. Significantly, the two seminal works on this material by N\'{e}el {\it et al.}~\cite{neel_magnetic_1964} and Paulev\'{e} {\it et al.}~\cite{pauleve_magnetization_1968} do indeed provide values for these separate $K_2$ and $K_3$ coefficients but they are unusually large for a uniaxial material. These large values of $K_2$ and $K_3$ are noted to make a significant contribution to the total uniaxial anisotropy coefficient, $K_U$.

In this work, we address the issues outlined above. We use a fully relativistic, first-principles-based framework to study the magnetocrystalline anisotropy of this system, providing results for both the equiatomic composition Fe$_{0.5}$Ni$_{0.5}$ as well as for an Fe-rich composition that is more consistent with experimentally determined composition of tetrataenite in some meteorite samples~\cite{clarke_tetrataeniteordered_1980, wasilewski_magnetic_1988, lewis_inspired_2014}, Fe$_{0.56}$Ni$_{0.44}$. We use our modelling framework to study the dependence of the magnetic torque on the direction of magnetisation and confirm that the leading-order anisotropy coefficient $K_1$ dominates higher-order coefficients $K_2$ and $K_3$ for the univariant single crystals modelled in this work. We go on to examine the magnetocrystalline anisotropy energy of this compound at finite temperature, and find that its value remains high up to and well beyond room temperature. We also consider the effect of varying long-range atomic order and find that, for both compositions, decreasing atomic order consistently decreases the computationally predicted MAE. { Finally, we perform calculations in which the effects of \emph{both} finite temperature and imperfect atomic ordering are included.} From a careful reading of the original analysis due to N\'{e}el {\it et al.}~\cite{neel_magnetic_1964}, we propose that our predicted MAE values for ideal single crystals are consistent with the experimental data, and that it is instead the nanocrystalline, multivariant nature of experimental samples~\cite{kovacs_discovery_2021} that yield larger values of $K_2$ and $K_3$ than expected {\it a priori}.

These results have significant implications. First, they provide insight into the discrepancies between measured and predicted anisotropies of this compound in the existing literature, emphasising that the higher-order anisotropy coefficients $K_2$ and $K_3$ appear to make a significant contribution to this material's experimentally observed total uniaxial anisotropy energy. Second, these results demonstrate the impact of varying Fe:Ni ratio on the predicted value of $K_1$, and conclude that an equiatomic composition is expected to exhibit superior magnetic performance compared to those of off-stoichiometry compositions. Finally, it is shown that the finite-temperature magnetocrystalline anisotropy of this material is remarkably robust, lending support for its potential use in elevated temperature applications. This work therefore supports renewed contemplation of this compound as a suitable material for advanced magnets.

This paper is structured as follows. In Section \ref{sec:methodology}, we briefly outline the underlying theory and our methodology for obtaining the MAE from first principles using the torque method as well as provide relevant computational details. Then, in Section \ref{sec:results}, we provide results of our calculations and carefully set them in an experimental context. Finally, in Section \ref{sec:conclusions}, we summarise our results, venture an outlook on their implications, and suggest possible further work.

\section{Methodology}
\label{sec:methodology}

Our technique for modelling the magnetocrystalline anisotropy energy of a material is based on evaluations of torque when a system is magnetised in a given direction.  These torques are obtained by evaluating derivatives of an energy, where the energy is obtained {\it ab initio} via density functional theory (DFT) calculations in which relativistic effects, notably spin-orbit coupling, are included~\cite{wang_torque_1996}. The method is a numerically robust way of calculating MAE values which are on the scale of 1~meV/atom or smaller. Further, by considering torque as a function of angle, the method enables direct evaluation of the separate anisotropy coefficients $K_1$, $K_2$, and $K_3$, as well as the usual uniaxial anisotropy constant $K_U$.

We use the Korringa-Kohn-Rostoker (KKR) formulation of DFT and the coherent potential approximation (CPA)~\cite{faulkner_multiple_2018} to describe the effects of partial atomic order~\cite{staunton_temperature_2006}. In addition, the effects of magnetic fluctuations at finite temperature are included via the disordered local moment (DLM) picture~\cite{gyorffy_first-principles_1985}.

\subsection{Magnetocrystalline anisotropy}
The magnetocrystalline anisotropy energy describes the variation in energy of a ferromagnetic system due to a rotation of the magnetisation direction. For a ferromagnet with magnetisation direction $\hat{\mathbf{n}}$, the MAE can be expressed as \cite{staunton_temperature_2006}
\begin{equation}
    K = \sum_\gamma \kappa_\gamma g_\gamma (\hat{\mathbf{n}}),
\end{equation}
where $\kappa_\gamma$ denote MAE coefficients and $g_\gamma$ denote spherical harmonics fixing the orientation of $\hat{\mathbf{n}}$ with respect to the crystal axes. It is convenient to represent $\hat{\mathbf{n}}$ in spherical polar coordinates by $\hat{\mathbf{n}} = (\sin\theta\cos\phi, \sin\theta\sin\phi, \cos\theta)$, where $\theta$ and $\phi$ denote polar and azimuthal angles, respectively, in a coordinate frame specified by the crystal axes. 

In a tetragonal uniaxial ferromagnet, the MAE is known to be well-described by~\cite{landau_electrodynamics_2009}
\begin{equation}
    K = K_1 \sin^2 \theta + K_2 \sin^4 \theta + K_3 \sin^4\theta \cos 4\phi,
\end{equation}
where $K_1$, $K_2$, and $K_3$ are the conventional anisotropy coefficients. The total energy of the system is then written as
\begin{equation}
    {\mathcal{E}}(\hat{\mathbf{n}}) = {\mathcal{E}}_\text{iso} + K_1 \sin^2 \theta + K_2 \sin^4 \theta + K_3 \sin^4\theta \cos 4\phi,
    \label{eq:uniax_e}
\end{equation}
where $\mathcal{E}_\text{iso}$ represents the isotropic portion of the energy. The MAE coefficients $K_1$, $K_2$, and $K_3$ can be accessed from derivatives of this energy with respect to magnetisation directions. Given an energy of the form in Eq.~\ref{eq:uniax_e}, we have that
\begin{align}
    \frac{\partial {\mathcal{E}}}{\partial \theta} =& K_1 \sin 2\theta + 2 K_2 \sin 2\theta \sin^2 \theta \nonumber \\
    &+ 2 K_3 \sin 2 \theta \sin^2 \theta \cos 4\phi,     \label{eq:ederiv1}\\
    \frac{\partial {\mathcal{E}}}{\partial \phi} =&  - 4 K_3 \sin^4\theta \sin 4\phi,
    \label{eq:ederiv2}
\end{align}
and evaluation of these quantities at particular values of $\theta$ and $\phi$ leads directly to the coefficients. 

Moreover, the magnetocrystalline anisotropy energy difference between the system magnetised in the $\hat{\mathbf{z}}$ and  $\hat{\mathbf{x}}$ directions, 
\begin{equation}
    K_U := {\mathcal{E}}((1,0,0)) - {\mathcal{E}}((0,0,1)) = K_1 + K_2 + K_3,
\end{equation}
is given by the evaluation of the derivative, $\frac{\partial \mathcal{E}}{\partial \theta}$ at $\theta = \frac{\pi}{4}$ and $\phi=0$ directly, enabling us to obtain the MAE in a one-shot calculation.
In practice, for a uniaxial ferromagnet, one expects $K_1$ to dominate $K_2$ and $K_3$ \cite{landau_electrodynamics_2009}, and the anisotropy is typically written as 
\begin{equation}
    K = K_1 \sin^2 \theta.
\end{equation}

\subsection{Evaluating magnetic torques \emph{ab initio}}

To evaluate these derivatives and, therefore, the MAE coefficients, we use an expression from the KKR multiple scattering formalism for the magnetic torque~\cite{staunton_temperature_2004, staunton_temperature_2006, patrick_marmot_2022, ouazi_atomic-scale_2012}, 
\begin{equation}
    \mathbf{T}^{(\hat{\mathbf{n}})} = - \frac{\partial \mathcal{E}^{(\hat{\mathbf{n}})}}{\partial \hat{\mathbf{n}}},
\end{equation}
in which the effects of finite-temperature local moment fluctuations~\cite{gyorffy_first-principles_1985} can also be included.  Using the stationary properties of the relativistic DFT energy $\mathcal{E}^{(\hat{\mathbf{n}})}$ with respect to the charge and magnetisation densities, the derivation of the torque begins with the single-electron energy-sum part of this energy.  In terms of the integrated electronic density of states,  $N^{(\hat{\mathbf{n}})}(E)$, for a system magnetised along a direction $\hat{\mathbf{n}}$, this energy is written as
\begin{equation}
{\mathcal{E}}^{(\hat{\mathbf{n}})} = - \int^{E^{(\hat{\mathbf{n}})}_F} \,N^{(\hat{\mathbf{n})}}(E) \, \textrm{d}E.
\end{equation}
In multiple scattering theory the integrated density of states is expressed particularly succinctly using the Lloyd formula~\cite{lloyd_wave_1967,faulkner_multiple_2018}
\begin{equation}
N^{(\hat{\mathbf{n}})}(E) = 
N_0 (E) -\frac{1}{\pi
}\operatorname{Im}\ln\det \left(  \underline{\underline{t}}(
\hat{\mathbf{n}}; E)
^{-1}-\underline{\underline{G}}_{0}(E))
\right), 
\end{equation}
where $\underline{\underline{t}} (\hat{\mathbf{n}}; E)$ describes an array of single-site-scattering t-matrices (combined into a super matrix in both site and angular momentum space) and $\underline{\underline{G}}_{0}(E)$ specifies the structure constants which contain all the information on the spatial location of the scatterers~\cite{faulkner_multiple_2018}. 

At a site $i$, the t-matrix describing the scattering of an electron from an effective scalar potential and local magnetic field, aligned with the spin-polarisation of the electron density in that region and located in the unit cell surrounding the site, is obtained from the solution of the Dirac equation. This solution is found in a coordinate frame in which the z-axis of the local coordinate frame is aligned with $\hat{\mathbf{n}}$~\cite{strange_relativistic_1984,ebert_calculating_2011}. A simple transformation produces a t-matrix for the general coordinate frame,
\begin{equation}
\underline{t}_i (\hat{\mathbf{n}}; E) =
\underline{R} (\hat{\mathbf{n}})\,\underline{t}%
_{i} (\hat{\mathbf{z}}; E)  \underline
{R}(\hat{\mathbf{n}})^{+}
\end{equation}
where $\underline{R} (\hat{\mathbf{n}})= \exp  i \alpha_{\hat{\mathbf{m}}} (\hat{\mathbf{m}} \cdot \mathbf{\underline{J}})$, $\alpha_{\hat{\mathbf{m}}}$ is the angle of rotation about an axis $\hat{\mathbf{m}} = \frac{\hat{\mathbf{z}} \wedge \hat{\mathbf{n}}}{|\hat{\mathbf{z}} \wedge \hat{\mathbf{n}}|}$, and $\mathbf{\underline{J}}$ is the total angular momentum.  The torque quantity $T_{\alpha_{\hat{u}}}^{( 
\hat{\mathbf{n}})  }= -\frac{\partial {\mathcal{E}}^{(\hat{\mathbf{n}})}}{\partial \alpha_{\hat{\mathbf{u}}}}$, describing the variation of the total energy with respect to a rotation of the magnetisation about a general
axis $\hat{\mathbf{u}}$, is 
\begin{widetext}
\begin{equation}
T_{\alpha_{\hat{\mathbf{u}}}}^{(\hat{\mathbf{n}})}= -\frac{1}{\pi} \int^{E^{( 
\hat{\mathbf{n}})}_F} \operatorname{Im}
\frac{\partial}{\partial \alpha_{\hat{\mathbf{u}}}} \left[ \ln\det\left(  \underline{\underline{t}} (
\hat{\mathbf{n}} ; E)^{-1}-\underline{\underline{G}}_{0}(E)
\right) \right] \textrm{d}E
\end{equation}
which can be written as \cite{staunton_temperature_2006}
\begin{equation}
T_{\alpha_{\hat{\mathbf{u}}}}^{(\hat{\mathbf{n}})}=
-\frac{1}{\pi} \int^{E^{( 
\hat{\mathbf{n}})}_F} \operatorname{Im} \sum_i \text{Tr}
\left( \underline{\tau}^{(\hat{\mathbf{n}})}_{ii} (E)
\frac{\partial}{\partial \alpha_{\hat{\mathbf{u}}}} \left( \underline{R}(\hat{\mathbf{n}}) \underline{t}(\hat{\mathbf{n}} ; E)^{-1} \underline{R}(\hat{\mathbf{n}} )^{+} \right)
\right) \textrm{d}E
\end{equation}
where the KKR scattering path operator~\cite{gyorffy_band_1973, faulkner_multiple_2018} is
\begin{equation}
\underline{\underline{\tau}}^{(\hat{\mathbf{n}})  }=\left(
\left(  \underline{\underline{t}}^{(\hat{\mathbf{n}})  }\right)
^{-1}-\underline{\underline{G}}_{0}\right)  ^{-1}\;. 
\end{equation}
Since $\frac{\partial \underline{R}(\hat{\mathbf{n}} 
)}{\partial \alpha_{\hat{\mathbf{u}}}} = i (\mathbf{\underline{J}} 
\cdot \hat{\mathbf{u}}) \underline{R}(\hat{\mathbf{n}})$ and 
$\frac{\partial \underline{R}(\hat{n})^{+}}{\partial 
\alpha_{\hat{\mathbf{u}}}} = - i (\mathbf{\underline{J}} \cdot \hat{\mathbf{u}}) 
\underline{R}(\hat{\mathbf{n}})$, we obtain
\begin{equation}
T_{\alpha_{\hat{\mathbf{u}}}}^{(\hat{\mathbf{n}})  }=\frac{1}{\pi} \int^{E^{( 
\hat{\mathbf{n}})}_F}  \operatorname{Im} 
i \, \sum_i
\text{Tr} \left( \underline{\tau}^{( \hat{\mathbf{n}})}_{ii}
(E)  \left[ 
(\mathbf{\underline{J}} \cdot \hat{\mathbf{u}})\, \underline{t}(\hat{\mathbf{n}} ; E
)^{-1} - \underline{t}(\hat{\mathbf{n}} ; E
)  ^{-1} (\mathbf{\underline{J}} \cdot \hat{\mathbf{u}}) \right] \right) \textrm{d}E. \label{T0}
\end{equation}
\end{widetext}
This closed-form expression enables us to evaluate the torque for a given magnetisation direction $\hat{\mathbf{n}}$. By sampling a range of values of $\theta$ and $\phi$, it is therefore possible to extract values of the coefficients $K_1$, $K_2$, and $K_3$ separately along with the conventional uniaxial anisotropy, $K_U$. We note that this technique for evaluating the MAE has previously exhibited exceptional fidelity when applied to other alloys crystallising in the L1\textsubscript{0} structure, giving values in excellent agreement with experimental literature for the FePd~\cite{staunton_temperature_2006} and FePt~\cite{staunton_temperature_2004} systems, along with explaining the dependence of the MAE on the long-range atomic order parameter~\cite{staunton_long-range_2004}. In addition, the method has also been used with success to study a number of rare-earth-based systems\cite{patrick_temperature-dependent_2019, bouaziz_crucial_2023} and other magnetic materials\cite{patrick_quantifying_2022}.

\subsection{Computational Details}
\label{sec:computational_details}

We use the all-electron HUTSEPOT code~\cite{hoffmann_magnetic_2020} to generate self-consistent, one-electron potentials within the KKR formulation of density functional theory (DFT)~\cite{martin_electronic_2004, faulkner_multiple_2018}. We perform spin-polarised, scalar-relativistic calculations within the atomic sphere approximation (ASA)~\cite{stocks_complete_1978} with an angular momentum cutoff of $l_\text{max} = 3$ for basis set expansions, a $20\times20\times20$ $\mathbf{k}$-point mesh for integrals over the Brillouin zone, and employ a 24-point semi-circular Gauss-Legendre grid in the complex plane to integrate over valence energies. We use the local density approximation (LDA) and the exchange-correlation functional of Perdew-Wang~\cite{perdew_accurate_1992}. For all calculations, we use lattice parameters of $a=b=3.560$ \AA, and $c=3.577$ \AA, obtained in an earlier first-principles study~\cite{izardar_interplay_2020}. The $c/a$ ratio is 1.0048, consistent with a very low tetragonality. These lattice parameters are broadly consistent with both earlier experimental~\cite{albertsen_tetragonal_1981, kotsugi_structural_2014, makino_artificially_2015, montes-arango_discovery_2016} and computational~\cite{miura_origin_2013, lewis_inspired_2014, edstrom_electronic_2014, werwinski_ab_2017} studies. The conventional simple tetragonal (A6) representation of the L1\textsubscript{0} unit cell with a two atom basis is used, as visualised in Fig.~\ref{fig:l10_structure}.

{ A comment should be made about the suitability of the ASA as used in this work to study the MAE via calculations of the magnetic torque of the FeNi system. The ASA description of the self-consistent potentials of the KKR-CPA formulation of DFT is known to be most suitable for description of close-packed systems~\cite{hoshino_accuracy_2001, ebert_calculating_2011}. L1\textsubscript{0} FeNi, with its underlying face-centred tetragonal lattice and a $c/a$ ratio close to unity, can be classified as close-packed and we therefore believe the ASA to be appropriate. In addition, the relativistic DLM picture required to study finite-temperature effects has only been implemented within a spherical potential approximation, such as the ASA employed in this work. So-called `full potential' schemes, such as those used in Ref.~\citenum{werwinski_ab_2017}, are expected to be necessary to accurately describe more complex crystal structures, especially if the MAE is calculated from direct energy differences rather than the magnetic torque.}

Utilization of the inhomogeneous CPA~\cite{faulkner_calculating_1980, johnson_total-energy_1990} allows observation of the effects of partial atomic order on the predicted MAE. An order parameter $S$ is defined to describe partial Fe-Ni atomic order in our configurations, where $S=1$ describes a maximally ordered state, and $S=0$ corresponds to a state which is completely disordered and both lattice sites shown in Fig.~\ref{fig:l10_structure} are equivalent. For the equiatomic composition, Fe$_{0.5}$Ni$_{0.5}$, the concentration of Fe on the 1a (1d) site is given by $0.5+0.5S$ ($0.5-0.5S$). For the Fe-rich composition, Fe$_{0.56}$Ni$_{0.44}$, the concentration of Fe on the 1a (1d) site is given by $0.56+0.44S$ ($0.56-0.44S$). The Ni concentrations can be inferred as the concentrations of Fe and Ni must sum to unity on all lattice sites.

To compute the MAE of our considered structures, we use the MARMOT code~\cite{patrick_marmot_2022}, which solves the single-site scattering problem on a fully relativistic footing, naturally including spin-orbit effects. The crystal structure for the MARMOT calculations is the same as that used for the HUTSEPOT calculations, and the angular momentum expansion was again truncated at $l_\text{max}=3$, while the remaining input parameters were left at the MARMOT default values. {This includes a $240\times40$ mesh for angular sampling of the CPA integral and an adaptive scheme for Brilluoin zone integrations.} MARMOT is used for both the zero and finite temperature calculations.

\section{Results}
\label{sec:results}

\subsection{Functional Form of the Anisotropy: $K_1$, $K_2$, and $K_3$}

To explore the functional form of the magnetocrystalline anisotropy we evaluate the magnetic torque over a range of values of $\theta$ and $\phi$ and fit the {\it ab initio} data to the form  given in Eqs.~\ref{eq:ederiv1} and \ref{eq:ederiv2}. Displayed in Table \ref{table:computational} are our obtained values of $K_1$, $K_2$, and $K_3$ for Fe$_{0.5}$Ni$_{0.5}$ and Fe$_{0.56}$Ni$_{0.44}$ at $T=0$ K with the perfect atomic order parameter $S=1.0$. Notably $K_1$ is almost four orders of magnitude larger than $K_2$ and $K_3$. We find similar results for calculations performed on systems with imperfect atomic order and at finite temperature; $K_1$ is always at least two orders of magnitude greater than $K_2$ and $K_3$. {We note that Ref~\citenum{werwinski_ab_2017} found a similar outcome in their computational study for perfectly ordered ($S=1$) L1$_0$ Fe$_{0.5}$Ni$_{0.5}$ at $T=0$ K; the leading-order anisotropy constant $K_1$ dominated higher-order constants $K_2$ and $K_3$. We are able to extend this result and find that the dominance of $K_1$ over $K_2$ and $K_3$ holds for systems modelled with imperfect atomic order and at finite temperature.} This outcome confirms that the MAE values of ideal single crystals modelled in our calculations are well-described by the standard formula for a uniaxial ferromagnet and, for the remainder of this paper, the torque will always be evaluated at the positions $\theta=\pi/4$, $\phi=0$ to directly obtain the uniaxial anisotropy constant $K_1 = K_U$.

\begin{table}[t]
\begin{ruledtabular}
\begin{tabular}{lrrrr}
Composition & \multicolumn{4}{l}{Anisotropy Constants (MJm$^{-3}$)} \\ \cline{2-5}
& $K_1$ & $K_2$ & $K_3$ & $K_U$\\ \hline
Fe$_{0.5}$Ni$_{0.5}$              & 0.9582       & $-$0.0008       & 0.0001       &  0.96       \\
Fe$_{0.56}$Ni$_{0.44}$            & 0.6122       & $-$0.0003       & 0.0006       &  0.61      
\end{tabular}
\end{ruledtabular}
\caption{Computed values of $K_1$, $K_2$, $K_3$, and $K_U$ for Fe$_{0.5}$Ni$_{0.5}$ and Fe$_{0.56}$Ni$_{0.44}$ with atomic order parameter $S=1$ and temperature $T=0$ K. It can be seen that $K_1$ is orders of magnitude larger than $K_2$ or $K_3$ for both compositions.}
\label{table:computational}
\end{table}

We calculate $K_1$ for maximally ordered (S=1) L1\textsubscript{0} Fe$_{0.5}$Ni$_{0.5}$ at $T=0$ K to be 0.96 MJm$^{-3}$, while for Fe$_{0.56}$Ni$_{0.44}$ we calculate it to be 0.61 MJm$^{-3}$. Our calculated MAE for the equiatomic Fe$_{0.5}$Ni$_{0.5}$ composition is broadly consistent with values of {0.23}--0.78~MJm$^{-3}$ that were previously obtained using a variety of other computational techniques~\cite{izardar_interplay_2020, si_effect_2022, werwinski_ab_2017, miura_origin_2013, tuvshin_first-principles_2021, edstrom_electronic_2014, lewis_magnete_2014}. However, we note that these values appear to be low in comparison with experimentally determined ones~\cite{neel_magnetic_1964, pauleve_magnetization_1968, shima_structure_2007, mizuguchi_artificial_2011, kojima_l1_2011, kojima_magnetic_2012, kotsugi_structural_2014, kojima_feni_2014, makino_artificially_2015, poirier_intrinsic_2015, frisk_resonant_2016, frisk_strain_2017, ito_fabrication_2023} of 0.2--0.93~MJm$^{-3}$, especially given that these measurements were performed on samples with imperfect atomic order and at room temperature. As noted in the introduction, analysis of the experimental data~\cite{neel_magnetic_1964, lewis_inspired_2014, lewis_magnete_2014} has suggested that the magnetocrystalline anisotropy energy of L1\textsubscript{0}-type FeNi could reach the range 1.0-1.3~MJm$^{-3}$.  Since {\it ab initio} computational estimates of the MAE for L1\textsubscript{0} transition metal systems are usually in good agreement with experimental data and, if anything, tend to overestimate the anisotropy constants, this computational underestimate stands out.

\subsection{Effects of finite temperature and imperfect atomic order}

To make the comparison between theory and experiment more direct we incorporate the effects of temperature and imperfect atomic order into the computational modelling. We can thus simulate the MAE of a single-variant FeNi crystal at room temperature with the atomic order parameter as furnished in the experimental reports (Table \ref{table:experimental}). 

\begin{figure*}[t]
\centering
\includegraphics[width=\textwidth]{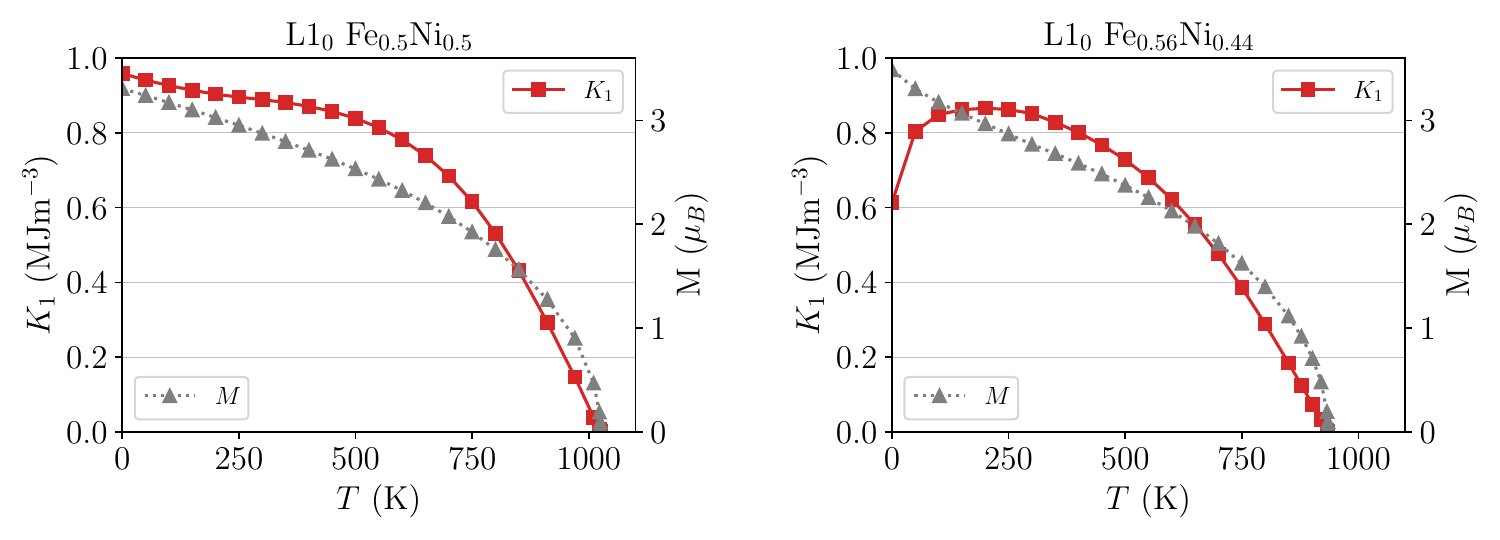}
\caption{Anisotropy constant $K_1$ and magnetisation $M$ calculated as a function of temperature for a perfectly ordered crystal of L1\textsubscript{0} Fe$_{0.5}$Ni$_{0.5}$ (left) {and for a maximally ordered crystal of L1\textsubscript{0} Fe$_{0.56}$Ni$_{0.44}$ (right). It can be seen that, in both cases, the predicted finite-temperature anisotropy remains high up to and beyond room temperature. Neither case obeys a simple power law of $K_1 \propto M^\alpha$. Notably, for the Fe-rich composition, the excess Fe on the Ni sublattice magnetically disorders very rapidly, indicating that the magnetocrystalline anisotropy of this composition at $T=300$ K is actually \emph{higher} than its zero-temperature value.}}
\label{fig:finite_t_anisotropy}
\end{figure*}

For these finite-temperature calculations, we use the disordered local moment (DLM) description of the magnetocrystalline anisotropy at finite temperature, as outlined in Sec.\ref{sec:methodology}. This method has been used with fidelity in the past to study other ferromagnetic L1\textsubscript{0}-structured materials~\cite{staunton_temperature_2004, staunton_temperature_2006} as well as rare-earth-based permanent magnets\cite{patrick_temperature-dependent_2019, bouaziz_crucial_2023}. 

{Shown in in the left panel of Fig.}~\ref{fig:finite_t_anisotropy} is a plot of $K_1$ for maximally ordered L1\textsubscript{0} Fe$_{0.5}$Ni$_{0.5}$ as a function of temperature, along with a plot of the total magnetisation, $M$, of the two-atom unit cell. It can be seen that, as expected, the anisotropy decreases as temperature increases, but that the decrease is very gradual, with the anisotropy energy at $T=300$ K computed as 0.889 MJm$^{-3}$, around 90\% of the $T=0$ value. This robust finite-temperature performance is intriguing; it is clear that the anisotropy energy does not obey a simple power law ($K_1 \propto M^\alpha$) as it does in other L1\textsubscript{0} materials such as FePt~\cite{staunton_temperature_2004}. This aspect was studied by Yamashita and Sakuma~\cite{yamashita_first-principles_2022, yamashita_finite-temperature_2023}, who used an atomistic spin model to find similar unusual finite temperature behaviour in L1\textsubscript{0} FeNi.

{Proceeding, in the right panel of Fig. \ref{fig:finite_t_anisotropy} is a plot of the same quantities for the Fe-rich composition in its maximally ordered state, with the 1a site having an occupancy of pure Fe, while the 1d site has occupancy Ni$_{0.88}$Fe$_{0.12}$. As we increase the temperature away from zero, there is a sharp jump in $K_U$ and an associated decrease in $M$, resulting in a predicted $K_U$ value at finite temperature which is only marginally lower than for the equiatomic composition. This sharp increase in $K_U$ for the Fe-rich composition is associated with a rapid magnetic disordering of the excess Fe on the 1d lattice site in our calculations. This excess Fe, which makes a significant negative contribution to the total magnetic torque in the zero-temperature calculations, makes a much smaller contribution to the torque once it has magnetically disordered, resulting in an increase in the predicted $K_U$ value for this composition. The rapid magnetic disordering is understood by considering the underlying face-centered tetragonal lattice. Pure Fe on an fcc lattice with this lattice spacing exhibits competition between ferromagnetic and antiferromagnetic ordering~\cite{pinski_ferromagnetism_1986}, and in this system this competition is manifest by a rapid disordering of the excess Fe on the Ni sites.}

Our computed Curie temperature of L1\textsubscript{0} Fe$_{0.5}$Ni$_{0.5}$ is 1025 K, in reasonable agreement with experimental measurements on the material that measured a Curie temperature of at least 830 K~\cite{lewis_inspired_2014}. We note that our Curie temperature determination is obtained within a mean-field theory and is therefore expected to be an overestimate of its true value. The Curie temperature { of 1220~K obtained by Yamashita and Sakuma~\cite{yamashita_first-principles_2022, yamashita_finite-temperature_2023}, using a perturbative approach to treat spin-orbit coupling, is higher than the value we obtain. For the Fe-rich composition, a marginally reduced Curie temperature of 935~K is predicted within our model. For both compositions, these Curie temperatures are theoretical estimates, as the material is expected to atomically disorder below this temperature\cite{neel_magnetic_1964, dos_santos_kinetics_2015}.}

\begin{table}[b]
\begin{ruledtabular}
\begin{tabular}{lllll}
\multirow{2}{*}{Composition}         & \multicolumn{4}{l}{Magnetic Moment ($\mu_B$)} \\
                                     & Total & Fe ($1a$)      & Ni ($1d$)     & Fe($1d$)     \\ \hline
Fe$_{0.5}$Ni$_{0.5}$                 & 3.24  & 2.62           &  0.62         &              \\
Fe$_{0.56}$Ni$_{0.44}$               & 3.43  & 2.60           &  0.63         & 2.26         
\end{tabular}
\end{ruledtabular}
\caption{{Magnetic moments associated with Fe and Ni $\mathrm{L}1_0$ systems considered in this work, at $T=0$~K and with an atomic order parameter of $S=1$. We also give the total magnetisation of the two-atom unit cell. The Fe-rich composition has excess Fe sitting on the $1d$ site.}}
\label{table:moments}
\end{table}

{A comment should also be made about magnetic moments predicted in our modelling. For reference, we provide the magnetic moments at $T=0$~K for an atomic order parameter of $S=1$ for both of the studied compositions in Table~\ref{table:moments}. The total magnetisation of the system compares well with experimental data, for example with the magnetisation of just over 1.6~$\mu_B$/atom for an equiatomic composition measured in Ref.~\citenum{frisk_strain_2017}.}

\begin{figure}[t]
\centering
\includegraphics[width=0.49\textwidth]{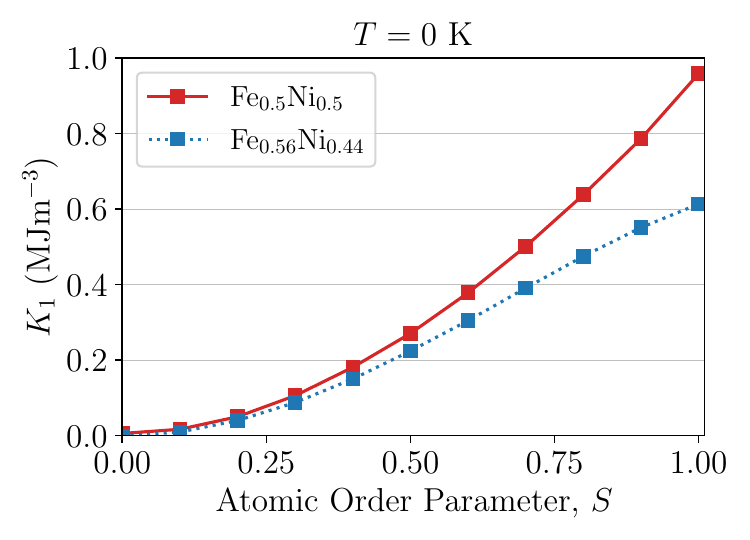}
\caption{MAE as a function of atomic long-range order parameter, $S$, for L1\textsubscript{0} Fe-Ni for both both the equatomic and Fe-rich compositions, { evaluated at $T=0$ K}. For both cases, decreasing long-range order decreases the predicted MAE, although the drop is sharper for the equiatomic composition than for the Fe-rich case.}
\label{fig:lro}
\end{figure} 

{We now move on to considering the effects of imperfect atomic ordering on the predicted MAE.} In Fig.~\ref{fig:lro}, we show the zero temperature MAE of both compositions as a function of atomic long-range order parameter, $S$. Decreasing atomic long-range order consistently decreases the calculated MAE for both compositions. This behavior is expected based on results obtained for other L1\textsubscript{0} compounds, such as FePt and FePd, from both experimental\cite{okamoto_chemical-order-dependent_2002} and computational\cite{staunton_long-range_2004} studies. We note that recent work by Izardar et al.\cite{izardar_interplay_2020}, which used supercell calculations to represent partially ordered L1\textsubscript{0} FeNi structures, concluded that the MAE for $S=0.75$ was comparable to that obtained from perfect $S=1.0$, with a value around 0.54~MJm$^{-3}$. This outcome is contrary to our results which show a sizeable MAE decrease with decreased atomic order. However, Izardar et al. utilized only 32 atoms in each supercell---a relatively small number---and obtained a large spread in calculated MAE values for the specific supercell configurations used to represent a particular order parameter.

\subsection{Comparison with experimental data}

\begin{figure}[b]
\centering
\includegraphics[width=0.49\textwidth]{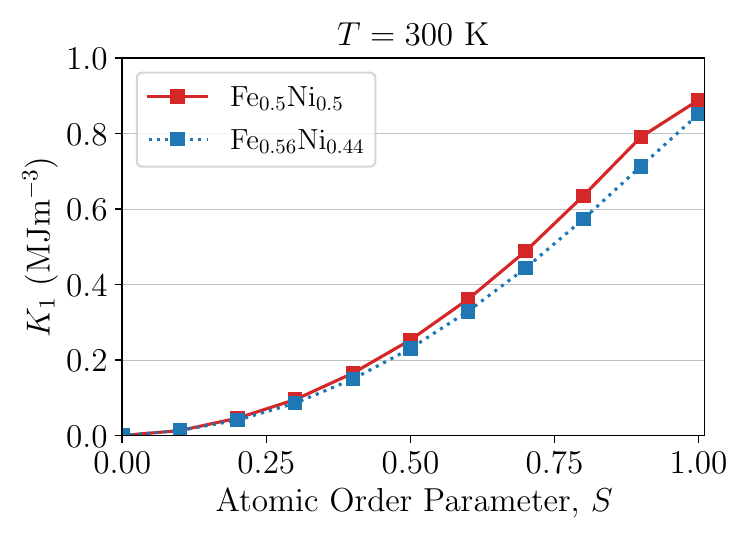}
    \caption{{ MAE as a function of atomic long-range order parameter, $S$, for L1\textsubscript{0} Fe-Ni for both both the equatomic and Fe-rich compositions, evaluated at $T=300$ K. As for the calculations at $T=0$~K, decreasing atomic order impacts the predicted anisotropy of the system. The predicted values of $K_1$ for Fe$_{0.5}$Ni$_{0.5}$ are only slightly decreased compared to their value at $T=0$~K, while for the Fe-rich composition the anisotropy at $T=300$~K is \emph{increased} compared to its zero temperature value, due to the rapid magnetic disordering of the excess Fe on the Ni sublattice.}}
\label{fig:lro_300K}
\end{figure}

{ To make as close as possible a comparison with the existing experimental data, we now provide results for calculations including the effects of \emph{both} imperfect atomic long-range order and finite temperature. Visualised in Fig.~\ref{fig:lro_300K} is a plot of $K_1$ against $S$ at a simulated temperature of $T=300$ K. When the results of the equiatomic composition, Fe$_{0.5}$Ni$_{0.5}$, are considered, it can be seen that $K_1$ is only fractionally decreased at 300~K compared to its value at 0~K, lending support for this material's potential use in applications with high operating temperatures. For the Fe-rich composition, there is a marked increase in $K_1$ across the range of $S$ values compared to the data obtained at $T=0$~K, particularly at larger values of $S$, where the system is close to being fully atomically ordered. This increase is associated with the unusual finite temperature behaviour visualised in the right-hand panel of Fig.~\ref{fig:finite_t_anisotropy}, associated with the excess Fe on the 1d lattice site rapidly disordering with increasing temperature.}

We now use the calculated temperature and atomic long-range order dependence of the MAE to make a direct comparison with the experimental values for FeNi summarised in Table~\ref{table:experimental}. { Considering first the total uniaxial anisotropy, $K_U$, it appears that the calculated values are significantly lower than the experimental ones. For example, the experimental determination of Ito {\it et al.}~\cite{ito_fabrication_2023} of $K_U = 0.55$~MJm$^{-3}$ for a single-variant thin film of L1\textsubscript{0} Fe$_{0.5}$Ni$_{0.5}$ with $S=0.6$ at $T=$ 293 K may be contrasted with with our calculated anisotropy coefficients, for the same conditions, of $K_1=0.3616$, $K_2=0.0005$, $K_3=-0.0001$~MJm$^{-3}$ that provide a uniaxial anisotropy constant of $K_U = 0.36$~MJm$^{-3}$, a markedly smaller value.}

Table~\ref{table:experimental}'s summary of experimental data is obtained from a variety of sources. We note that not all references disclose the atomic order parameter $S$ for the sample, making a direct comparison with our results challenging. In addition, only two references by Ne\'{e}l and co-workers~\cite{neel_magnetic_1964, pauleve_magnetization_1968} resolve the measured anisotropy, $K_U$ into its separate components, $K_1$, $K_2$, $K_3$, and of these two references only one~\cite{neel_magnetic_1964} provides an atomic long-range order parameter in addition to the measured anisotropy coefficients.

{ When we consider the resolution of the total uniaxial anisotropy into its separate coefficients, $K_1$, $K_2$, and $K_3$, the origin of the discrepancy becomes clear. For example, the} experimental value of $K_1$ = 0.32~MJm$^{-3}$ reported by N\'{e}el and coworkers in Ref.~\citenum{neel_magnetic_1964} for a specimen of estimated order parameter $S=0.41$--0.45 at $T=293$~K is in close alignment with our results, as shown in Fig.~\ref{fig:lro_300K}. However, Ref.~\citenum{neel_magnetic_1964} also reports experimental determinations of large values of $K_2$ and $K_3$, which we do not find.  These large values are unexpected for a uniaxial ferromagnetic material, an assessment noted by the authors of the original study. Indeed one of the authors, N\'{e}el, developed a theory to account for the unexpectedly large $K_2$ and $K_3$ values in these nanocrystalline samples~\cite{neel_magnetic_1964, pauleve_magnetization_1968}. 

The picture described by N\'{e}el is of a polycrystalline sample consisting of individual nanocrystals with L1\textsubscript{0} atomic layering directions (determining the magnetic easy axis) that are not all aligned along a common direction. (Note that there are three possible directional variants of the L1\textsubscript{0} structure, with layering perpendicular to $\hat{\mathbf{x}}$, $\hat{\mathbf{y}}$, or $\hat{\mathbf{z}}$ directions.) N\'{e}el proposes that these nanocrystals are smaller than the size of magnetic domains within the sample, with a suggested nanocrystal size of approximately 150 \AA. It is concluded that exchange coupling between nanocrystals can lead to finite macroscopic values of $K_2$ and $K_3$, even though individual nanocrystals themselves are assumed to have a negligible value of these coefficients. It is known that meteoritic tetrataenite samples have nanoscale structure~\cite{kovacs_discovery_2021}. In addition, laboratory-grown films often contain step edges, dislocations and other structural imperfections which may lead to similar exchange-coupling mechanisms that could impact the measured anisotropy. We emphasise that these large $K_2$ and $K_3$ values, which are not captured within first-principles calculations, appear to make a significant contribution to the total uniaxial anisotropy, $K_U$, of this system. 

N\'{e}el's argument is predicated on his assumption that the application of a magnetic field along the $\hat{\mathbf{x}}$ direction during the annealing process will favor L1\textsubscript{0} layering along that same direction. To quantify this effect, we use our computational formalism for modelling atomic arrangements in alloys\cite{woodgate_compositional_2022, woodgate_short-range_2023, woodgate_interplay_2023} in which relativistic effects and, therefore spin-orbit coupling, can be included \cite{staunton_temperature_2004}. We find that, for a sample magnetised along the $\hat{\mathbf{x}}$ direction, the predicted ordering is indeed an L1\textsubscript{0}-type layering along the $\hat{\mathbf{x}}$ direction, with an ordering temperature of 508 K. However, the difference in energy between layering along $\hat{\mathbf{x}}$ direction and layering along $\hat{\mathbf{y}}$ or $\hat{\mathbf{z}}$ directions is only  0.7 MJm$^{-3}$, comparable to our calculated magnetocrystalline anisotropy energy constant $K_1$. This small but detectable difference supports N\'{e}el's hypothesis that samples annealed with an applied field parallel to the $\hat{\mathbf{x}}$ direction will produce crystallites with L1\textsubscript{0} layering along all three crystal axes, but layering along the $\hat{\mathbf{x}}$ direction will be favoured over that along the other two directions. The hypothesis that an inhomogeneous, nanostructured sample can yield unexpectedly large higher-order anisotropy coefficients is also supported by other authors, for example, by recent computational work on inhomogeneous FePt using atomistic spin modelling~\cite{binh_higher-order_2023}.

We suggest, therefore, that more data from both experimental measurements and computational modelling are required to resolve this intriguing apparent discrepancy between theory and experiment. It may be that a mechanism, similar to that outlined by N\'{e}el, is unrecognised but at play in a number of the experimental datasets. Further experimental work that resolves the magnetocrystalline anisotropy into its separate $K_1$, $K_2$, and $K_3$ components will enable definitive conclusions to be drawn about this disparity.  In addition, atomistic and micromagnetic modelling are anticipated to provide further insight into the proposed exchange coupling mechanism and attendant magnetic properties of the multivariant geometry described by N\'{e}el.

\section{Conclusions}
\label{sec:conclusions}

In summary, we have carried out an {\it ab initio} computational study of the magnetocrystalline anisotropy energy of two L1\textsubscript{0}-type FeNi compositions, equiatomic Fe$_{0.5}$Ni$_{0.5}$ and Fe-rich Fe$_{0.56}$Ni$_{0.44}$. This rare-earth-free compound is of interest as a sustainable advanced permanent magnet material. We confirm that the modeled magnetic torque of ideal single crystals is well-described by the standard formula for a uniaxial ferromagnet, with the leading-order magnetocrystalline anisotropy energy coefficient $K_1$ dominating the higher-order coefficients $K_2$ and $K_3$. {This dominance persists to finite temperature and is also predicted in samples with imperfect atomic order.} Moreover we find that a significant MAE persists to high temperatures, affirming this material's potential as a useful `gap' magnet.

To compare computational outcomes as faithfully as possible with experimental ones, we have included the effects of both finite temperature and imperfect atomic order on the predicted magnetocrystalline anisotropy. It is confirmed that both increasing temperature and decreasing atomic order decrease the MAE of this compound. We have made a detailed comparison of our computed anisotropy coefficients with those of the experimental literature and find that our calculated values of $K_1$ appear to be consistent with the experimental data. However, the experimental reports from N\'eel {\it et al.}~\cite{neel_magnetic_1964} and Paulev\'e {\it et al.}~\cite{pauleve_magnetization_1968} detail significant $K_2$ and $K_3$ values, which we do not find.

We suggest that the noted discrepancy between computational and experimental literature values of the uniaxial magnetocrystalline anistropy constant $K_U$ has its origins in these large $K_2$ and $K_3$ values observed in the early experiments. These values significantly contribute to the experimentally observed uniaxial magnetocrystalline anisotropy constant, $K_U$. It is proposed that these large $K_2$ and $K_3$ values have their origins in the polycrystalline, multivariant nature of experimental samples.

Further experimental work should seek to resolve separate values for $K_1$, $K_2$, and $K_3$ at both very low and at finite temperatures to better understand the magnetocrystalline anisotropy of this material. Further computational/theoretical work could focus on micromagnetic or atomistic spin modelling to better understand the model proposed by N\'{e}el~\cite{neel_magnetic_1964} and its consequences.

\begin{acknowledgments}
We gratefully acknowledge the support of the UK Engineering and Physical Sciences Research Council, Grant No. EP/W021331/1. C.D.W. is supported by a studentship within the UK Engineering and Physical Sciences Research Council-supported Centre for Doctoral Training in Modelling of Heterogeneous Systems, Grant No. EP/S022848/1. This work was also supported in by the U.S. Department of Energy, Office of Basic Energy Sciences under Award Number DE SC0022168 (for atomic insight) and by the U.S. National Science Foundation under Award ID 2118164 (for advanced manufacturing aspects).
\end{acknowledgments}

\section*{Author Declarations}

\subsection*{Journal Submission}

This article has been submitted to the Journal of Applied Physics.

\subsection*{Conflict of Interest}

The authors have no conflicts to disclose.

\subsection*{Author Contributions}

{\bf Christopher D. Woodgate:} Conceptualization (equal); Data curation (lead); Formal analysis (lead); Investigation (lead); Methodology (equal); Validation (lead); Visualisation (lead); Writing -- original draft (lead); Writing -- review \& editing (lead). {\bf Christopher E. Patrick:} Software (supporting); Writing -- review \& editing (supporting). {\bf Laura H. Lewis:} Conceptualization (equal); Funding Acquisition (equal); Project administration (equal); Supervision (equal); Writing -- review \& editing (supporting). {\bf Julie B. Staunton:} Conceptualization (equal); Funding Acquisition (equal); Methodology (equal); Project administration (equal); Resources (lead); Supervision (equal); Writing -- review \& editing (supporting).

\section*{Data Availability Statement}

The data that support the findings of this study are available from the corresponding author upon reasonable request.

\end{document}